\documentstyle[epsf,12pt]{article}
\begin{document}
\begin{center}
\Large{\bf 2D Azimuthal Space for Au + Au Mid-Central Collisions at $\sqrt{s_{NN}} =$ 200 GeV}\\
\large{R.S. Longacre$^a$\\
$^a$Brookhaven National Laboratory, Upton, NY 11973, USA}
\end{center}
 
\begin{abstract}
In this paper we will show that one can summarize the major two particle 
reaction plane azimuthal correlations for Au + Au mid-central collisions at 
$\sqrt{s_{NN}} =$ 200 GeV by defining a 2D azimuthal space which is a summary
of the event by event average.
\end{abstract}
 
\section{Introduction to Three Particle and Reaction Plane Two Particle 
Azimuthal Correlations.} 

In this paper we are interested in four three particle azimuthal correlations.
These correlations are defined by the four equations below

\begin{equation}
C_{112} = \langle cos(\phi_1 + \phi_2 - 2\phi_3)\rangle ,
\end{equation}

\begin{equation}
C_{123} = \langle cos(\phi_1 + 2\phi_2 - 3\phi_3)\rangle ,
\end{equation}

\begin{equation}
C_{132} = \langle cos(\phi_1 - 3\phi_2 + 2\phi_3)\rangle ,
\end{equation}

\begin{equation}
C_{143} = \langle cos(\phi_1 - 4\phi_2 + 3\phi_3)\rangle ,
\end{equation}

where $\phi_1$, $\phi_2$, $\phi_3$ denote the azimuthal angles of the
produced particle 1, produced particle 2 and produced particle 3. 

For Au + Au mid-central collisions at $\sqrt{s_{NN}} =$ 200 GeV, the quadrupole
flow from an impact angle defines the orientation Au + Au collision and defined
the angle of $v_2$. This $v_2$ angle is charge independent only determined
by geometry. If we choose $\phi_3$ to be charge independent and coming from
particles which range over the whole event, we can replace $\phi_3$ with 
$\psi_2$ the quadrupole flow angle for the first and third equations. For each 
event there will also be a hexapole flow angle for $v_3$ which gives a $\psi_3$ 
which can be used in the second and fourth equations.

\clearpage

\begin{equation}
C_{112} = \langle cos(\phi_1 + \phi_2 - 2\psi_2)\rangle ,
\end{equation}

\begin{equation}
C_{123} = \langle cos(\phi_1 + 2\phi_2 - 3\psi_3)\rangle ,
\end{equation}

\begin{equation}
C_{132} = \langle cos(\phi_1 - 3\phi_2 + 2\psi_2)\rangle ,
\end{equation}

\begin{equation}
C_{143} = \langle cos(\phi_1 - 4\phi_2 + 3\psi_3)\rangle ,
\end{equation}

For this paper we will choose $\psi_2$ to be zero. This requires that all events
are rotated in azimuth such that $\psi_2$ becomes zero. In general the 
$\psi_3$ points in some other direction. With this choice of $\psi_2$, we
show in Ref\cite{mono} that there is a mono jet effect with squeeze out flow 
of particles around the mono jet which gives a $\psi_3$ pointing in the 
$-90^\circ$ or  -$\pi\over 2$ direction. This implies that the mono jet points 
in the $90^\circ$ or  $\pi\over 2$ direction.

Also this paper will consider particles produced around central rapidity and
concentrate on two particle azimuthal correlations(see above). There is an 
another two particle azimuthal correlation which does not depend on any axis
 
\begin{equation}
\delta =  \langle cos(\phi_1 - \phi_2)\rangle .
\end{equation}

There is a simple relationship between equation 5 and equation 9 since 
$\psi_2$\ = 0.. 

\begin{equation}
C_{112} = \langle cos(\phi_1 + \phi_2) \rangle = \langle cos(\phi_1)cos(\phi_2)
\rangle - \langle sin(\phi_1) sin(\phi_2) \rangle ,
\end{equation}

\begin{equation}
\delta = \langle cos(\phi_1 - \phi_2) \rangle = \langle cos(\phi_1) cos(\phi_2) 
\rangle + \langle sin(\phi_1) sin(\phi_2) \rangle  .
\end{equation}

Thus $\langle$ $cos(\phi_1)$ $cos(\phi_2)$ $\rangle$ and  
$\langle$ $sin(\phi_1)$ $sin(\phi_2)$ $\rangle$ are also values we calculate 
in this paper.
  .

Finally the quadrupole and the hexapole are global event objects while $\delta$
the two particle correlation is a more local effect. In the appendix we show
that just the presents of $v_2$, $v_3$ and a $\delta$ can generate three
particle correlation. In our analysis we are dealing with the Case II of the 
appendix which consider only particles produced around central rapidity and the
results from Au + Au mid-central collisions at $\sqrt{s_{NN}} =$ 200 GeV are
inconsistent with the appendix.

The paper is organized in the following manner:

Sec. 1 is the introduction to the two particle reaction plane azimuthal 
correlations for heavy ion Au + Au mid-central collisions. Sec. 2 the 
additions to Ref\cite{mono} same sign pairs correlation. Sec. 3 $\phi_1$ 
and $\phi_2$ azimuthal two particle plane. Sec. 4 the two particle correlation 
from the $\phi_1$ and $\phi_2$ azimuthal plane. Sec. 5 calculation of 
correlations using the distribution of the $\phi_1$ and $\phi_2$ azimuthal 
plane. Sec. 6 presents the summary and discussion.

\section{Additions to Ref\cite{mono} Same Sign Pairs Correlation.} 

In this paper we where inspired by the squeeze out of particles around the mono 
jet of Ref\cite{mono} to establish an orientation of the hexapole flow axis
with respect to the quadrupole flow axis. In the introduction we determined
that $\langle$ $cos(\phi_1)$ $cos(\phi_2)$ $\rangle$ and  
$\langle$ $sin(\phi_1)$ $sin(\phi_2)$ $\rangle$ are very important
correlations to be explored. Thus we see in Ref\cite{mono} when we compare
to Ref\cite{Csepvsplane} there is a systematic difference. In the flux tube
model\cite{QGP,tubevsjet} there is conservation of momentum between the
flux tubes such that there is a negative correlation when one compares particles
coming from tubes at different pseudo rapidity($\eta$). When we concentrate
on like sign particle pairs this effect is much stronger than the model of 
Ref\cite{mono}. 

In Figure 1 we show the $\langle$ $cos(\phi_1)$ $cos(\phi_2)$ $\rangle$ and  
$\langle$ $sin(\phi_1)$ $sin(\phi_2)$ $\rangle$ for same sign pairs and 
opposite sign pairs as a function $\Delta \eta$($\vert \eta_1 - \eta_2 \vert$) 
of Ref\cite{mono}. This should be compared to FIG 8 of Ref\cite{Csepvsplane}. 
For like sign particle pairs between a flux tube and other tubes there is a 
greater a negative correlation of momentum conservation at smaller 
$\Delta \eta$ than Ref\cite{mono} so we increase this negative correlation 
to become in agreement with Ref\cite{Csepvsplane} see Figure 2. 
With the addition of this same negative correlation to the like sign particle 
pairs correlation for  $\langle$ $cos(\phi_1)$ $cos(\phi_2)$ $\rangle$ and  
$\langle$ $sin(\phi_1)$ $sin(\phi_2)$ $\rangle$ we do not cause a change in
the results of Ref\cite{mono}. This is because they depend on the difference
between the two terms and thus subtract out.

\begin{figure}
\begin{center}
\mbox{
   \epsfysize 8.0in
   \epsfbox{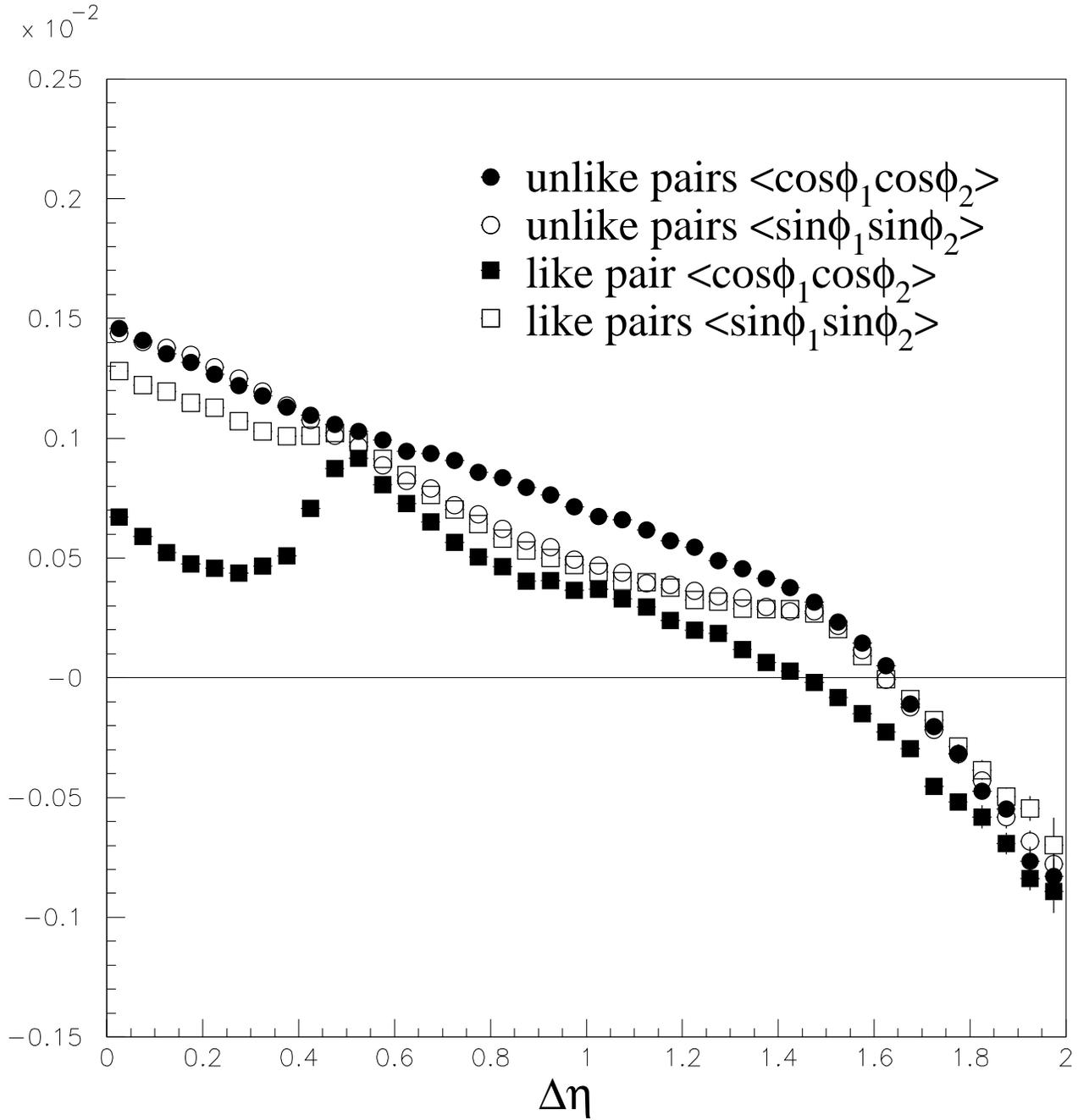}}
\end{center}
\vspace{2pt}
\caption{ An angular correlation 
$\langle cos(\phi_1(\eta_1) + \phi_2(\eta_2)) \rangle$ 
for like sign and unlike sign pairs is split up into  
in plane $\langle cos(\phi_1(\eta_1)) cos(\phi_2(\eta_2)) \rangle$ and out of 
plane $ \langle sin(\phi_1(\eta_1)) sin(\phi_2(\eta_2)) \rangle$ parts. The 
in plane is the flux tubes while the out of plane is the CME(like sign) and
the mono jet effect(both). The simulation is for mid-central Au-Au collision 
$\sqrt{s_{NN}}$ = 200.0 GeV}
\label{fig1}
\end{figure}

\begin{figure}
\begin{center}
\mbox{
   \epsfysize 8.0in
   \epsfbox{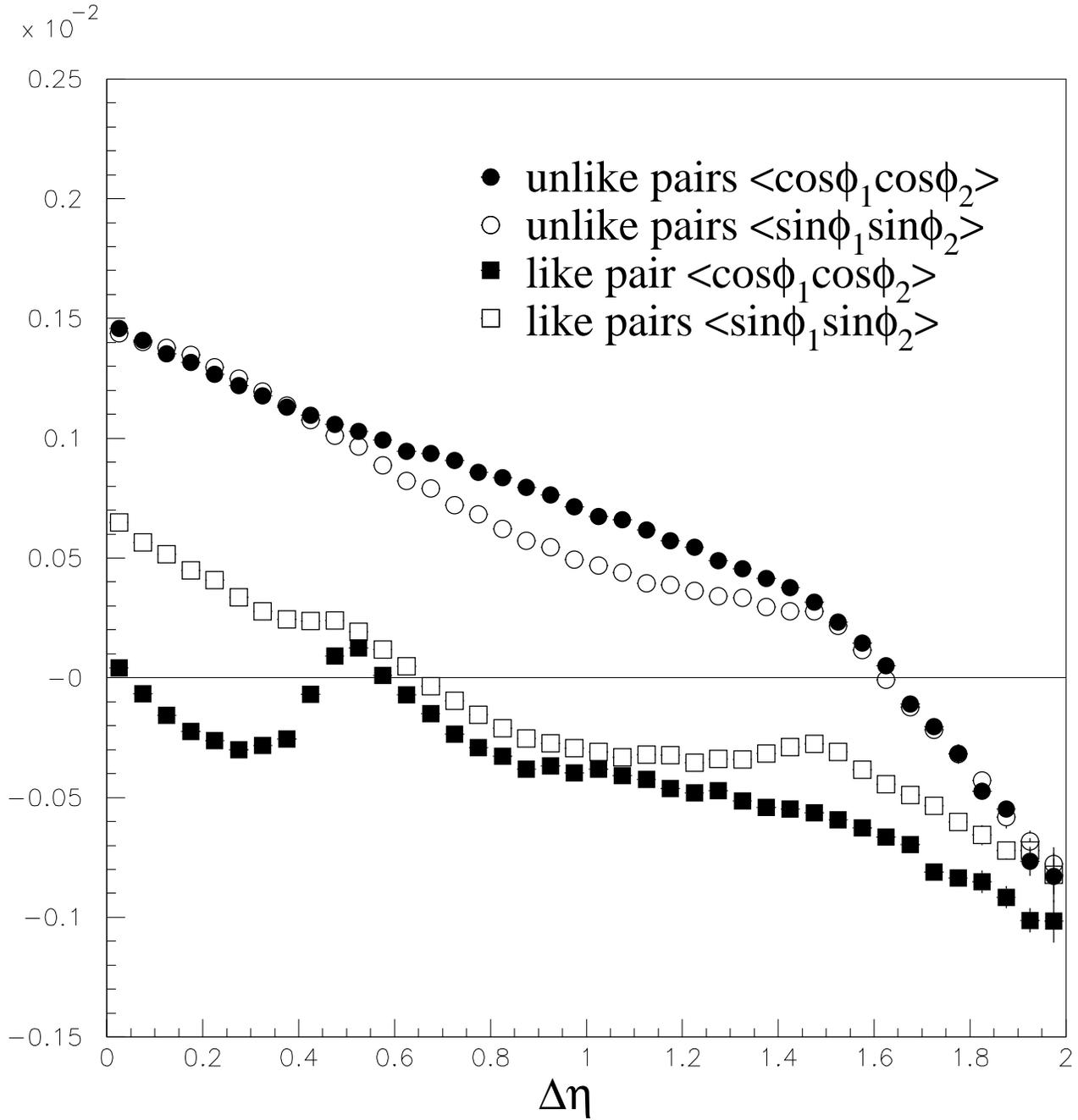}}
\end{center}
\vspace{2pt}
\caption{ Like sign and unlike sign pairs as seen in Figure 1 is split up into  
in plane $\langle cos(\phi_1(\eta_1)) cos(\phi_2(\eta_2)) \rangle$ and out of 
plane $ \langle sin(\phi_1(\eta_1)) sin(\phi_2(\eta_2)) \rangle$ parts. 
A negative correlation of momentum conservation is add to both 
in plane and out of plane like sign pairs only so it agrees with 
Ref\cite{Csepvsplane}. The simulation is for mid-central Au-Au collision 
$\sqrt{s_{NN}}$ = 200.0 GeV}
\label{fig2}
\end{figure}

\begin{figure}
\begin{center}
\mbox{
   \epsfysize 6.0in
   \epsfbox{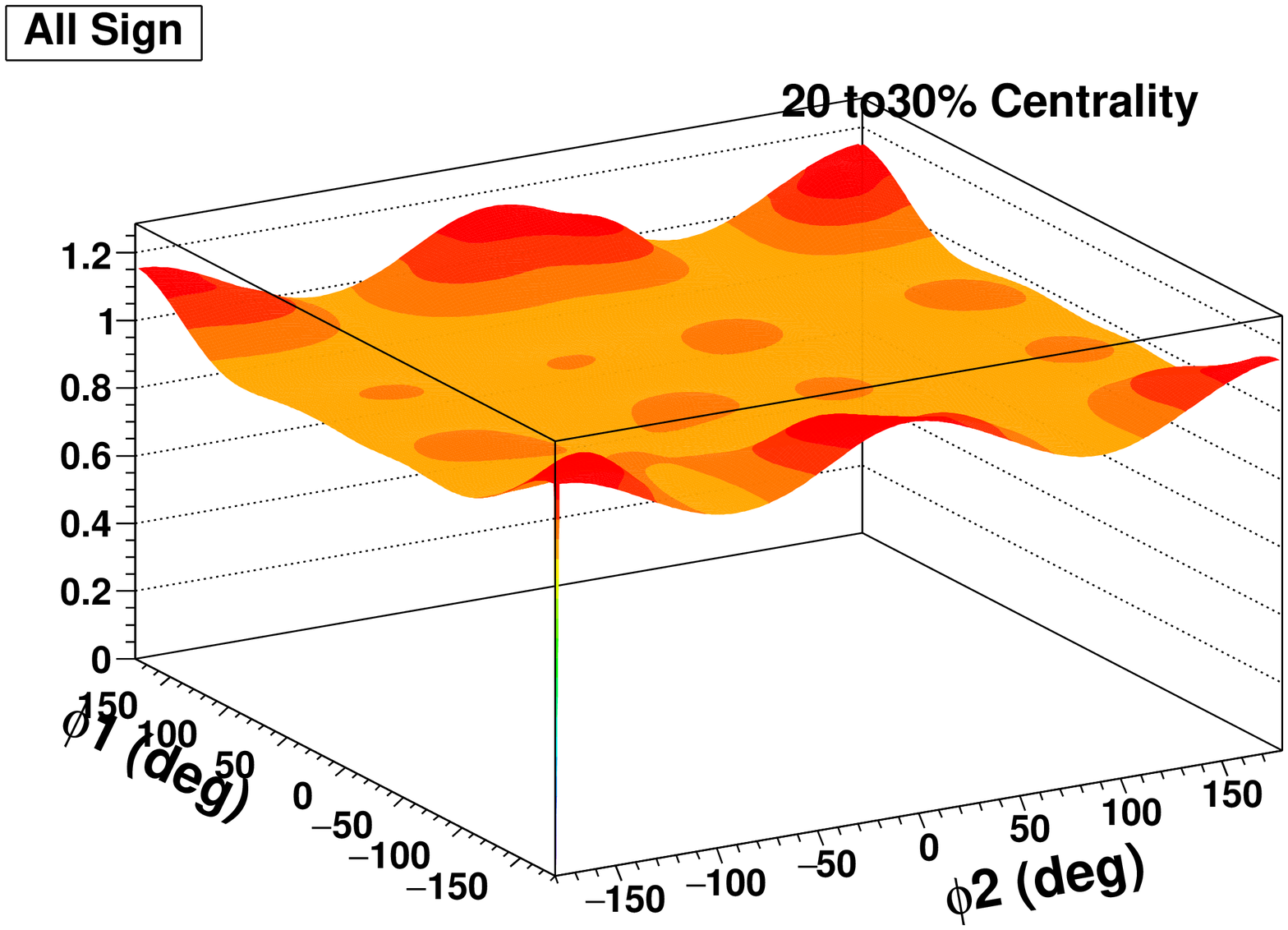}}
\end{center}
\vspace{2pt}
\caption{The reaction plane has two azimuthal angles $\phi_1$ and $\phi_2$ where
geometry defined an $v_2$ axis to lie in the $0^\circ$ to $180^\circ$ axis. 
In Ref\cite{mono} the presence of out of plane mono jets then gives us a $v_1$ 
which we require to point along the $90^\circ$ axis. The squeeze out flow 
around the hot spot of mono jet causes the $v_3$ axis to point along the 
$-90^\circ$ axis. We show the distribution for all particle pairs.}
\label{fig3}
\end{figure}

\section{$\phi_1$ and $\phi_2$ Azimuthal Two Particle Plane.}

The main objective of this paper is to introduce the concept of summarizing
the azimuthal two particle correlations with respect to the reaction plane
for heavy ion Au + Au mid-central collisions, by using a $\phi_1$ and $\phi_2$ 
azimuthal two particle plane. For each mid-central collision we have a well
defined $v_2$ axis given by geometry. We assign this well defined axis to lie 
in the $0^\circ$ to $180^\circ$ axis. In Ref\cite{mono} the presence of out of 
plane mono jets then gives us a $v_1$ which we require to point along the 
$90^\circ$ axis. The squeeze out flow around the hot spot of mono jet causes 
the $v_3$ axis to point along the $-90^\circ$ axis.

This paper contends that the event average of the azimuthal two particle angles
$\phi_1$ and $\phi_2$ with respect to the reaction plane as defined above
captures the azimuthal correlation structure of the Au + Au mid-central 
collisions. We use Ref\cite{mono} modified as pointed out in section 2 to
generate a event average $\phi_1$ and $\phi_2$ plane distribution. In
Figure 3 we show the event averaged $\phi_1$ and $\phi_2$ plane distribution
for all particle pairs which we have generated from the above method.

\clearpage

In Figure 4 we show the event averaged $\phi_1$ and $\phi_2$ plane distribution
for opposite charged particle pairs. Opposite charges are greatly influenced by
the fragmentation of the charge neutral quark gluon plasma. We see an asymmetry
in the $\phi_1$ and $\phi_2$ plane distribution, because once we choose one
charge the distribution of the other charge is different due to fragmentation
effects.

Finally in Figure 5 we show the event averaged $\phi_1$ and $\phi_2$ plane 
distribution for same charged particle pairs. The same charges must have a 
symmetry in the $\phi_1$ and $\phi_2$ plane, since we have Bose symmetry 
particle 1 and 2 are interchangeable.

\begin{figure}
\begin{center}
\mbox{
   \epsfysize 6.0in
   \epsfbox{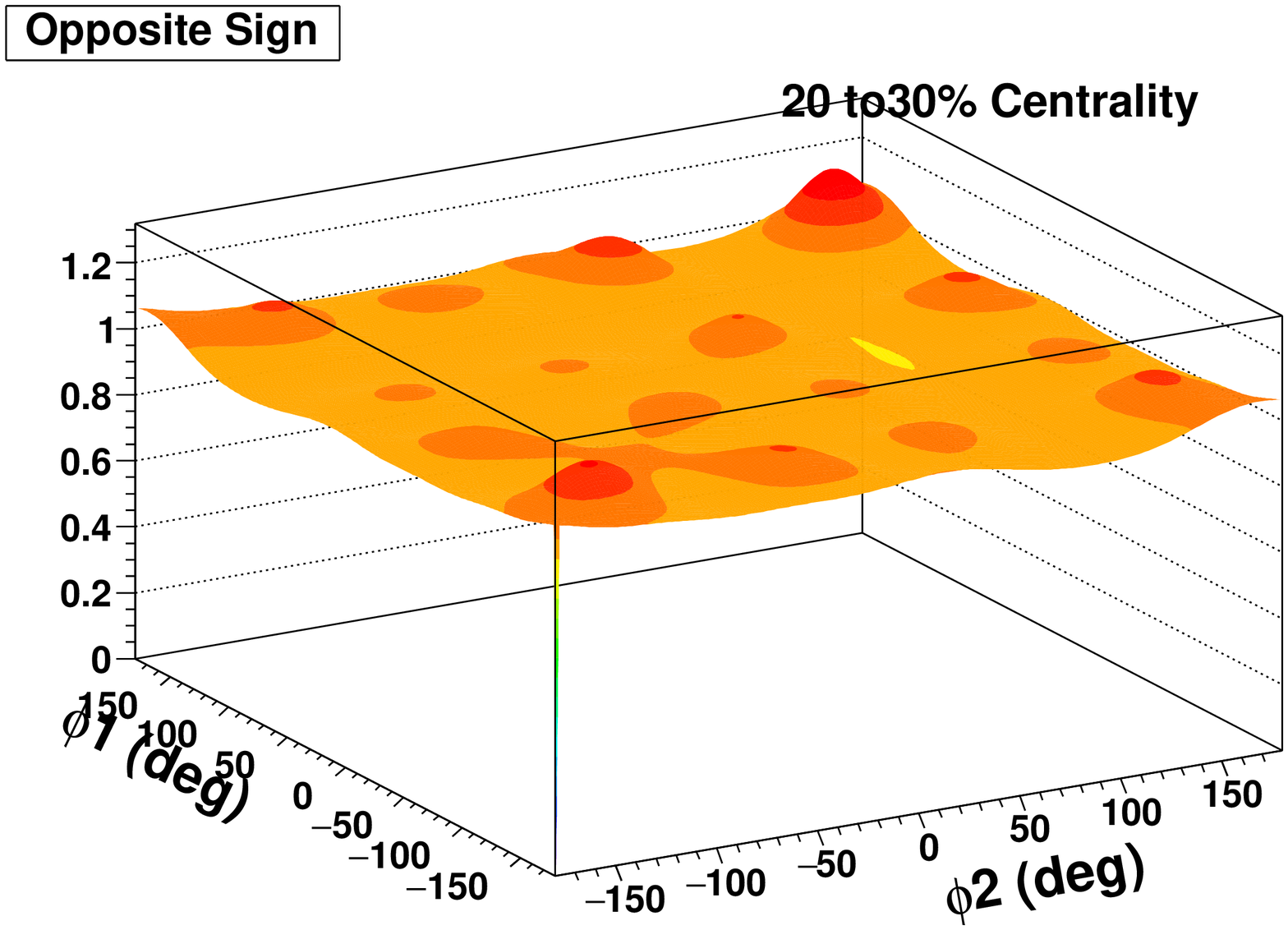}}
\end{center}
\vspace{2pt}
\caption{The reaction plane has two azimuthal angles $\phi_1$ and $\phi_2$ where
geometry defined an $v_2$ axis to lie in the $0^\circ$ to $180^\circ$ axis. 
In Ref\cite{mono} the presence of out of plane mono jets then gives us a $v_1$ 
which we require to point along the $90^\circ$ axis. The squeeze out flow 
around the hot spot of mono jet causes the $v_3$ axis to point along the 
$-90^\circ$ axis. We show the distribution for unlike sign particle pairs.}
\label{fig4}
\end{figure}

\begin{figure}
\begin{center}
\mbox{
   \epsfysize 6.0in
   \epsfbox{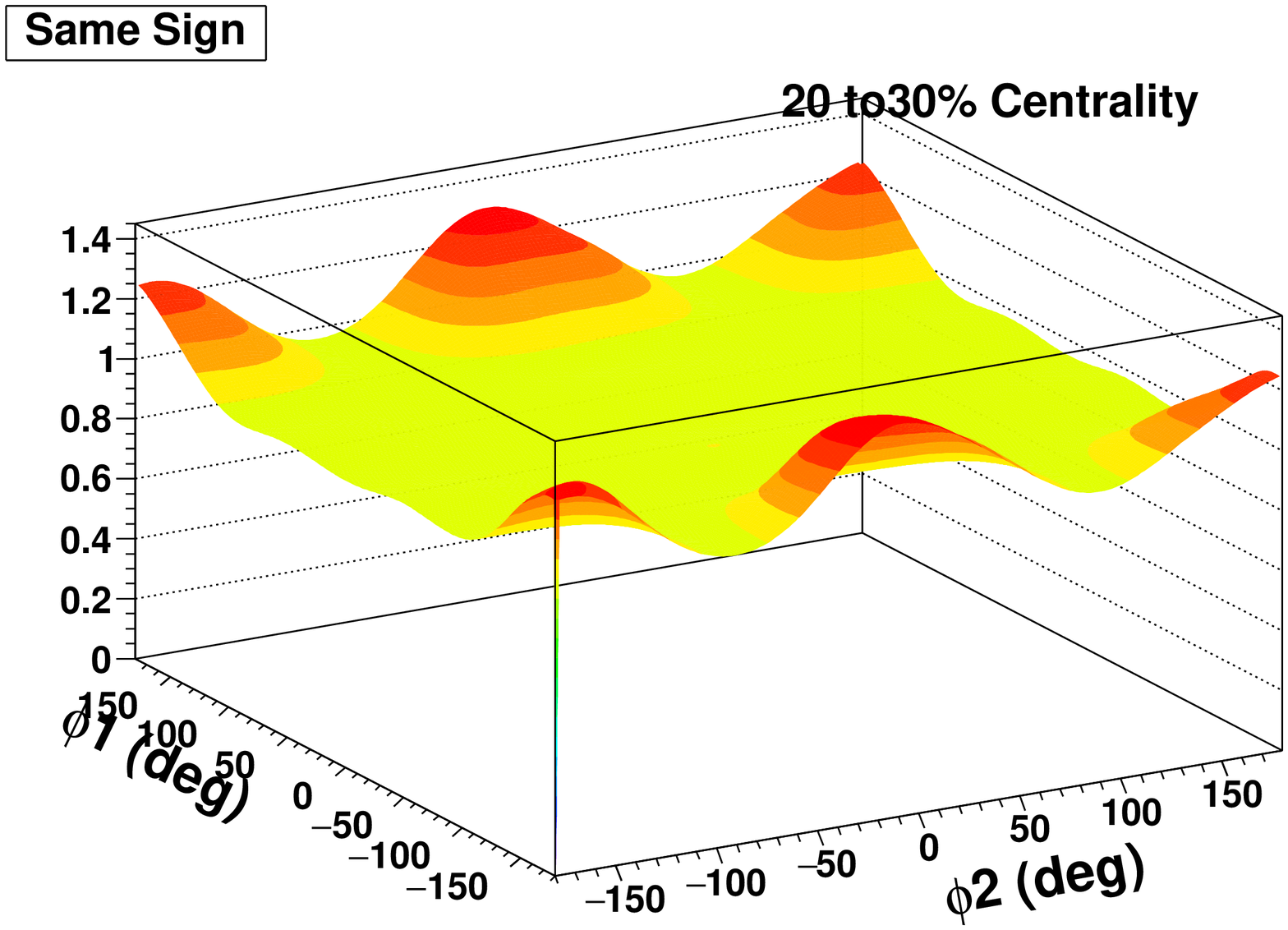}}
\end{center}
\vspace{2pt}
\caption{The reaction plane has two azimuthal angles $\phi_1$ and $\phi_2$ where
geometry defined an $v_2$ axis to lie in the $0^\circ$ to $180^\circ$ axis. 
In Ref\cite{mono} the presence of out of plane mono jets then gives us a $v_1$ 
which we require to point along the $90^\circ$ axis. The squeeze out flow 
around the hot spot of mono jet causes the $v_3$ axis to point along the 
$-90^\circ$ axis. We show the distribution for like sign particle pairs.}
\label{fig5}
\end{figure}

\begin{figure}
\begin{center}
\mbox{
   \epsfysize 7.5in
   \epsfbox{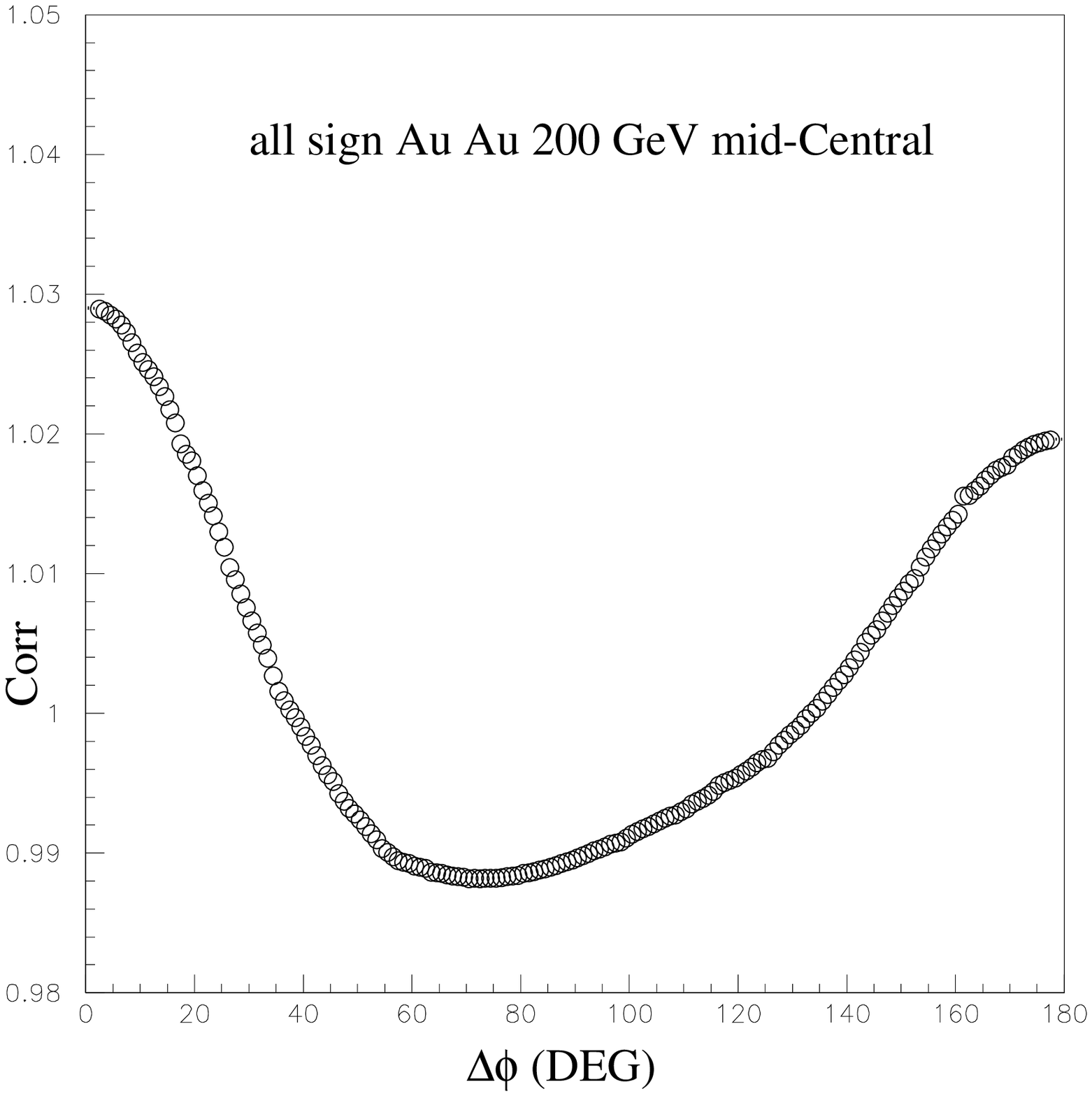}}
\end{center}
\vspace{2pt}
\caption{The reaction plane has two azimuthal angles $\phi_1$ and $\phi_2$ where
geometry defined an $v_2$ axis to lie in the $0^\circ$ to $180^\circ$ axis. 
In Ref\cite{mono} the presence of out of plane mono jets then gives us a $v_1$ 
which we require to point along the $90^\circ$ axis. The squeeze out flow 
around the hot spot of mono jet causes the $v_3$ axis to point along the 
$-90^\circ$ axis. The two particle correlation 
$\langle cos(\phi_1 - \phi_2) \rangle$ is shown for all particle pairs.}
\label{fig6}
\end{figure}

\begin{figure}
\begin{center}
\mbox{
   \epsfysize 7.5in
   \epsfbox{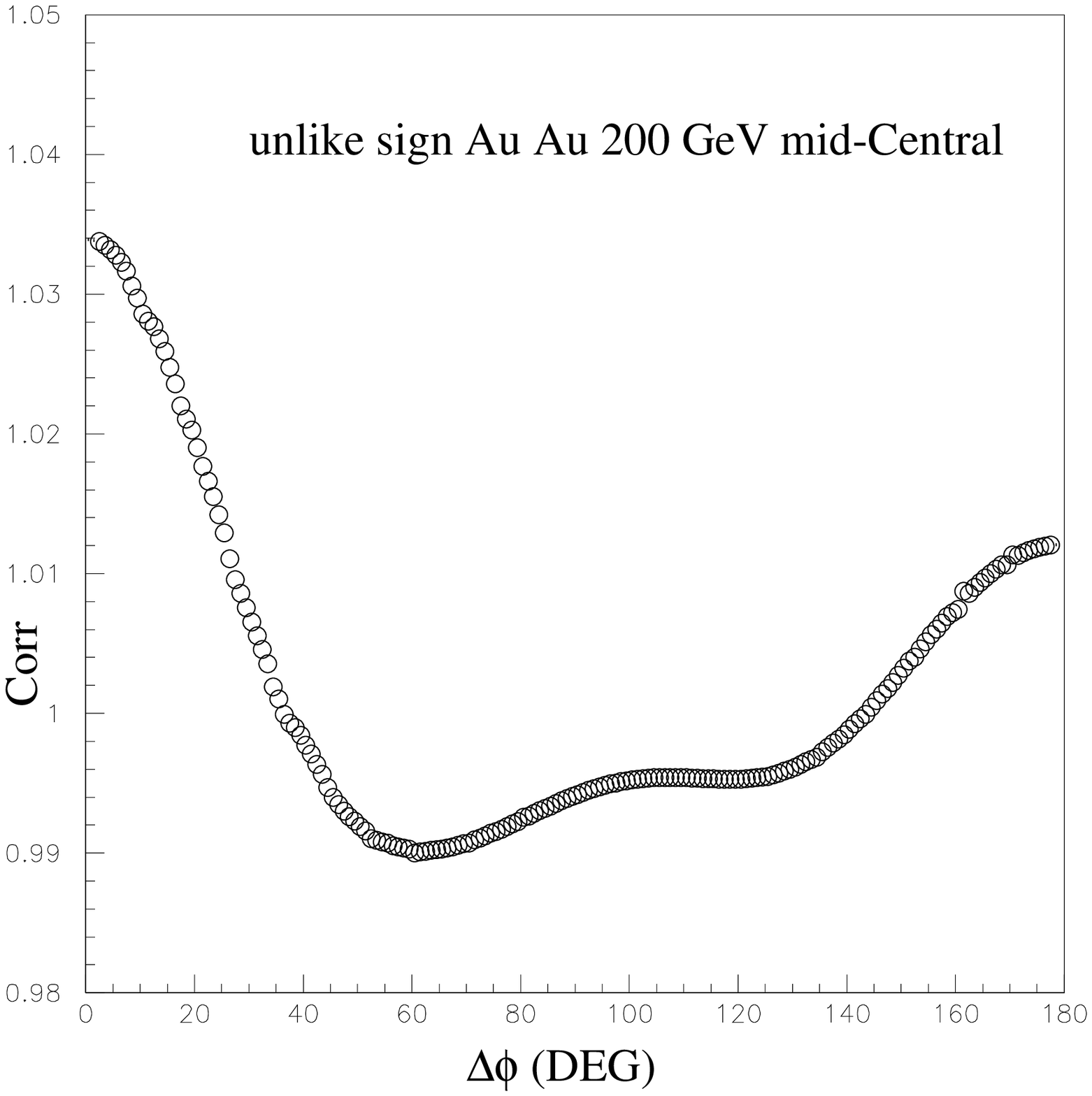}}
\end{center}
\vspace{2pt}
\caption{The reaction plane has two azimuthal angles $\phi_1$ and $\phi_2$ where
geometry defined an $v_2$ axis to lie in the $0^\circ$ to $180^\circ$ axis. 
In Ref\cite{mono} the presence of out of plane mono jets then gives us a $v_1$ 
which we require to point along the $90^\circ$ axis. The squeeze out flow 
around the hot spot of mono jet causes the $v_3$ axis to point along the 
$-90^\circ$ axis. The two particle correlation 
$\langle cos(\phi_1 - \phi_2) \rangle$ is shown for unlike sign particle pairs.}
\label{fig7}
\end{figure}

\begin{figure}
\begin{center}
\mbox{
   \epsfysize 7.5in
   \epsfbox{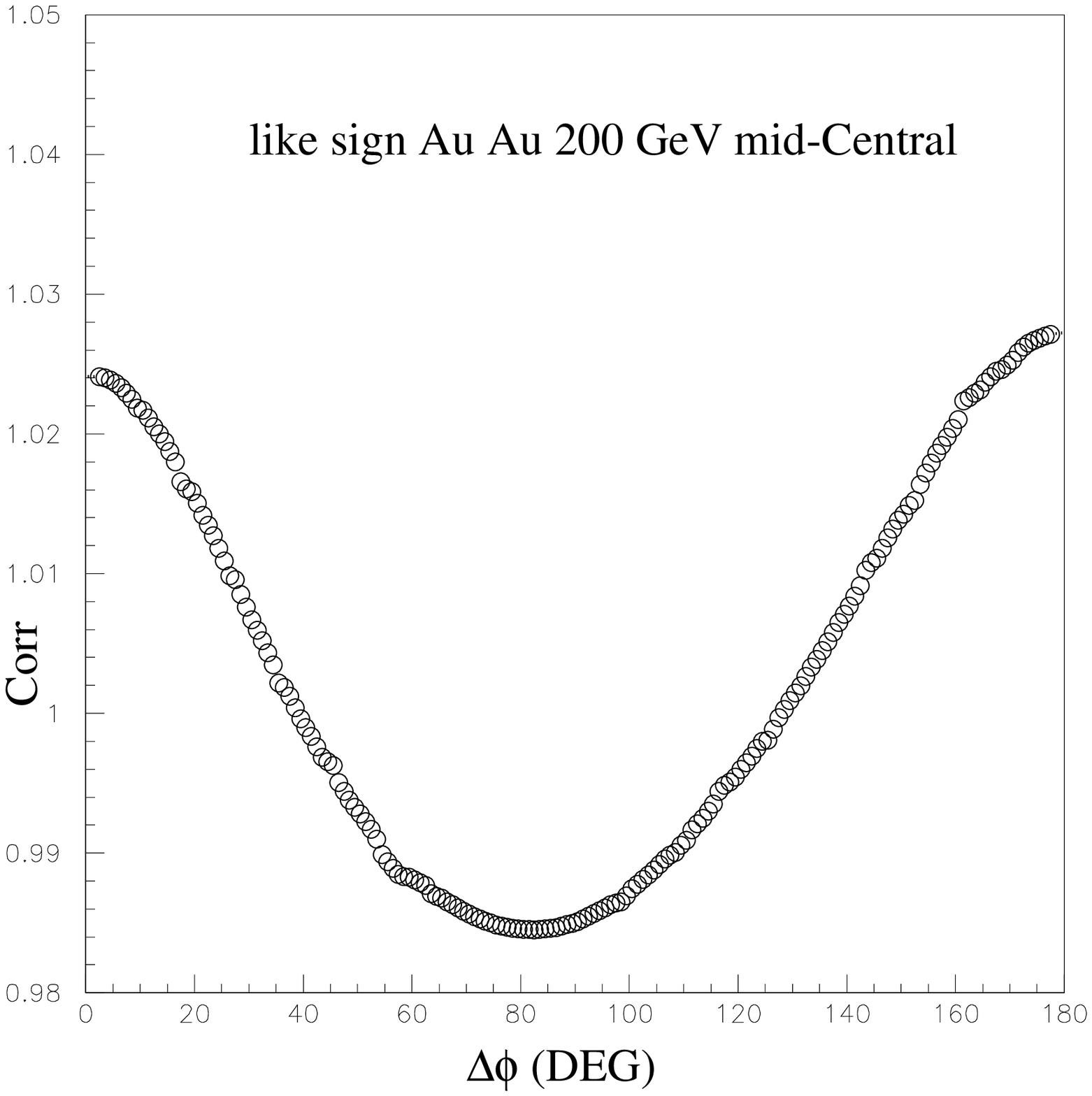}}
\end{center}
\vspace{2pt}
\caption{The reaction plane has two azimuthal angles $\phi_1$ and $\phi_2$ where
geometry defined an $v_2$ axis to lie in the $0^\circ$ to $180^\circ$ axis. 
In Ref\cite{mono} the presence of out of plane mono jets then gives us a $v_1$ 
which we require to point along the $90^\circ$ axis. The squeeze out flow 
around the hot spot of mono jet causes the $v_3$ axis to point along the 
$-90^\circ$ axis. The two particle correlation 
$\langle cos(\phi_1 - \phi_2) \rangle$ is shown for like sign particle pairs.}
\label{fig8}
\end{figure}

\section{The Two Particle Correlation from the $\phi_1$ and $\phi_2$ Azimuthal 
Plane.}

The two particle azimuthal correlation($\delta$) for all charged particle paairs
is defined by

\begin{equation}
\delta = \langle cos(\phi_1 - \phi_2) \rangle,
\end{equation}

which is independent of the reaction plane. this is because if any angle is 
added to the azimuth it will be subtracted away in the difference. In Figure 6
we plot $\delta$ for all charge particle pairs as a function $\Delta \phi$. 
$\Delta \phi$ is given by

\begin{equation}
\Delta \phi = (\phi_1 - \phi_2).
\end{equation}

Figure 6 show the presence of a dipole and a quadrupole. In Figure 7 we 
plot $\delta$ for unlike sign particle pairs as a function $\Delta \phi$. 
Figure 7 show the presence of a dipole, a quadrupole and a hexapole. In Figure 
8 we plot $\delta$ for like sign particle pairs as a function $\Delta \phi$. 
Figure 8 shows the presence of a negative dipole and a quadrupole. 

\section{Calculation of Correlations using the Distribution of the $\phi_1$ and 
$\phi_2$ Azimuthal Plane}

In this section we calculate the correlations introduce in section 1. This 
includes $C_{112}$, $C_{123}$, $C_{132}$, and $C_{143}$. We also calculate
$\langle$ $cos(\phi_1)$ $cos(\phi_2)$ $\rangle$ and  
$\langle$ $sin(\phi_1)$ $sin(\phi_2)$ $\rangle$. For the calculation of
$C_{112}$(equation 5) and $C_{132}$(equation 7) we use  $\psi_2$ = 0 because
the quadrupole flow angle in $\phi_1$ and $\phi_2$ azimuthal plane is in the
the $0^\circ$ to $180^\circ$ axis. For the calculation of $C_{123}$(equation 6) 
and $C_{143}$(equation 8) we use  $\psi_3$ = $-90^\circ$ axis because
the hexapole flow angle in $\phi_1$ and $\phi_2$ azimuthal plane is caused
by squeeze out flow around the hot spot of mono jets which is the 
$-90^\circ$ axis. Finally we calculate $\langle$ $cos(\phi_1)$ 
$cos(\phi_2)$ $\rangle$ and  $\langle$ $sin(\phi_1)$ $sin(\phi_2)$ $\rangle$.
The results of these calculations for unlike sign pairs are shown in Table I
and Figure 9, where as the calculations for like sign pairs are shown in 
Table II and Figure 10. 

\clearpage

\begin{figure}
\begin{center}
\mbox{
   \epsfysize 3.0in
   \epsfbox{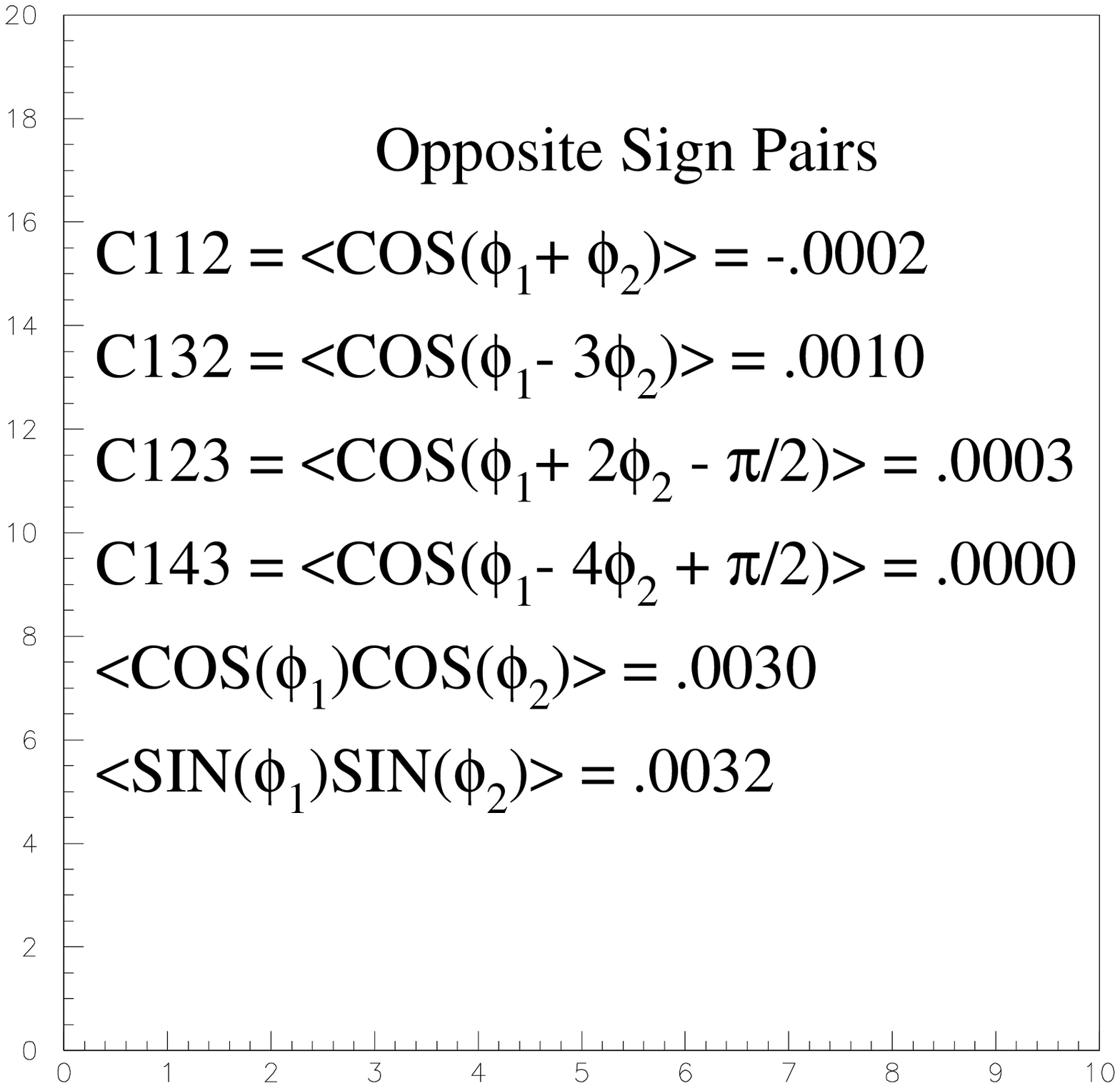}}
\end{center}
\vspace{2pt}
\caption{ Unlike Sign.}
\label{fig9}
\end{figure}

\clearpage

\begin{figure}
\begin{center}
\mbox{
   \epsfysize 3.0in
   \epsfbox{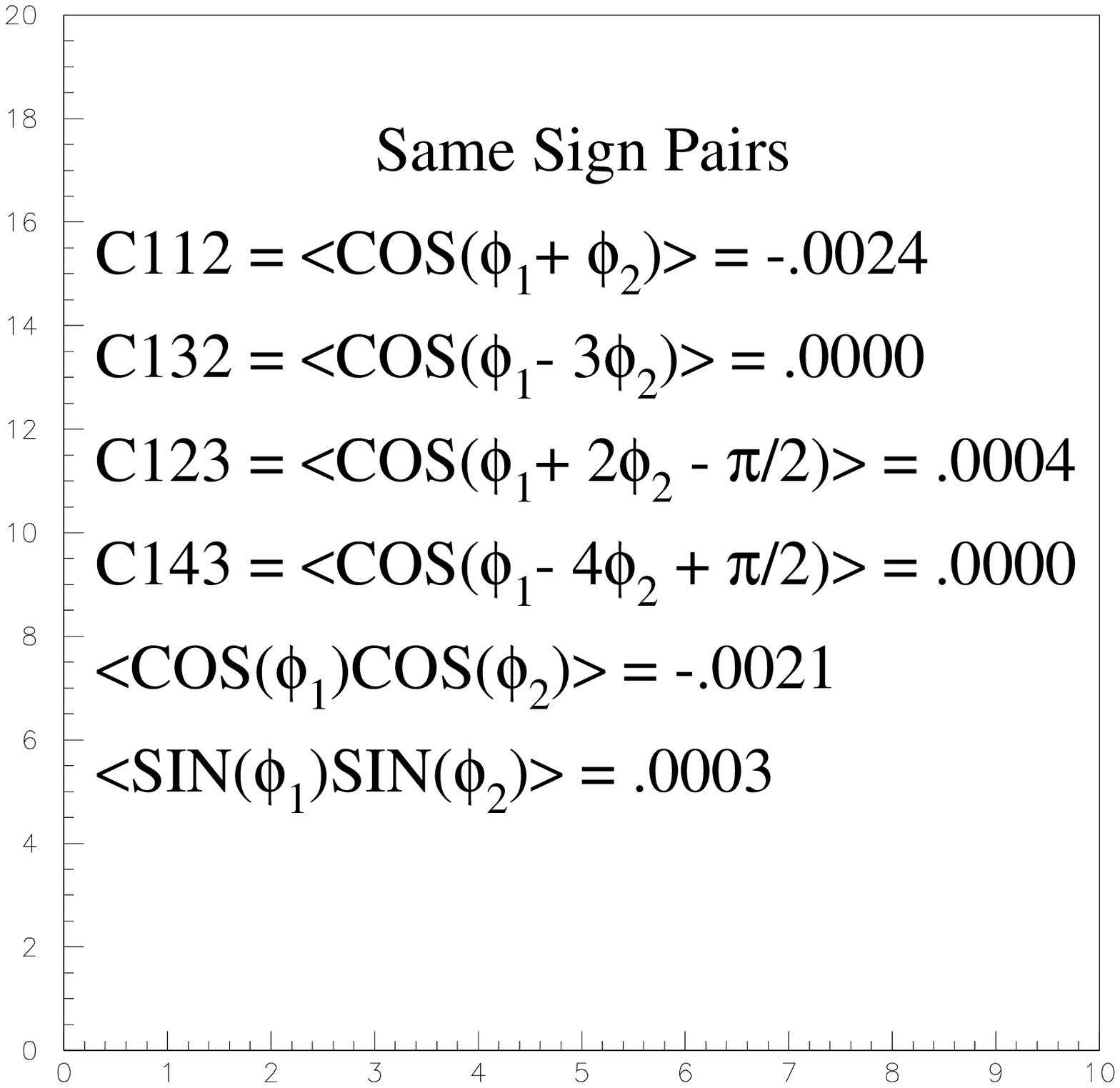}}
\end{center}
\vspace{2pt}
\caption{ Like Sign.}
\label{fig10}
\end{figure}
\clearpage

\bf Table I. \rm Correlations for Unlike Sign Pairs.

\begin{center}
\begin{tabular}{|c|r|}\hline
\multicolumn{2}{|c|}{Table I}\\ \hline
correlation & unlike sign pairs\\ \hline
C112 = $<$cos($\phi_1$+$\phi_2$)$>$ & -.0002 \\ \hline
C132 = $<$cos($\phi_1$-3$\phi_2$)$>$ & .0010 \\ \hline
C123 = $<$cos($\phi_1$+2$\phi_2$-${\pi\over 2}$)$>$ & .0003\\ \hline
C143 = $<$cos($\phi_1$-4$\phi_2$+${\pi\over 2}$)$>$ & .0000\\ \hline
$<$cos($\phi_1$)cos($\phi_2$)$>$ & .0030\\ \hline
$<$sin($\phi_1$)sin($\phi_2$)$>$ & .0032\\ \hline
\end{tabular}
\end{center}

\bf Table II. \rm Correlations for Like Sign Pairs.

\begin{center}
\begin{tabular}{|c|r|}\hline
\multicolumn{2}{|c|}{Table II}\\ \hline
correlation & like sign pairs\\ \hline
C112 = $<$cos($\phi_1$+$\phi_2$)$>$ & -.0024 \\ \hline
C132 = $<$cos($\phi_1$-3$\phi_2$)$>$ & .0000 \\ \hline
C123 = $<$cos($\phi_1$+2$\phi_2$-${\pi\over 2}$)$>$ & .0004\\ \hline
C143 = $<$cos($\phi_1$-4$\phi_2$+${\pi\over 2}$)$>$ & .0000\\ \hline
$<$cos($\phi_1$)cos($\phi_2$)$>$ & -.0021\\ \hline
$<$sin($\phi_1$)sin($\phi_2$)$>$ & .0003\\ \hline
\end{tabular}
\end{center}

\section{Summary and Discussion}

The main objective of this paper is to introduce the concept of summarizing
the azimuthal two particle correlations with respect to the reaction plane
for heavy ion Au + Au mid-central collisions, by using a $\phi_1$ and $\phi_2$ 
azimuthal two particle plane. For each mid-central collision we have a well
defined $v_2$ axis given by geometry. We assign this well defined axis to lie 
in the $0^\circ$ to $180^\circ$ axis. In Ref\cite{mono} the presence of out of 
plane mono jets then gives us a $v_1$ which points along the $90^\circ$ axis. 
The squeeze out flow around the hot spot of mono jet causes the $v_3$ axis 
to point along the $-90^\circ$ axis.

We use  the squeeze out of particles around the mono jet of Ref\cite{mono} 
to establish an orientation of the hexapole flow axis with respect to the 
quadrupole flow axis. In section 1 we determined that 
$\langle$ $cos(\phi_1)$ $cos(\phi_2)$ $\rangle$ and  
$\langle$ $sin(\phi_1)$ $sin(\phi_2)$ $\rangle$ are very important
correlations to be explored. Thus we see in Ref\cite{mono} when we compare
to Ref\cite{Csepvsplane} there is a systematic difference. In the flux tube
model\cite{QGP,tubevsjet} there is conservation of momentum between the
flux tubes such that there is a negative correlation when one compares particles
coming from tubes at different pseudo rapidity($\eta$). When we concentrate
on like sign particle pairs this effect is seen to be stronger than the model 
of Ref\cite{mono}. In Figure 1 we show the $\langle$ $cos(\phi_1)$ 
$cos(\phi_2)$ $\rangle$ and  $\langle$ $sin(\phi_1)$ $sin(\phi_2)$ $\rangle$ 
for same sign pairs and opposite sign pairs as a function 
$\Delta \eta$($\vert \eta_1 - \eta_2 \vert$) of Ref\cite{mono}. This should 
be compared to FIG 8 of Ref\cite{Csepvsplane}. For like sign particle pairs 
between a flux tube and other tubes there is a greater a negative correlation 
of momentum conservation at smaller $\Delta \eta$ than Ref\cite{mono} so we 
increase this negative correlation to become in agreement with 
Ref\cite{Csepvsplane} see Figure 2. With the addition of this same negative 
correlation to the like sign particle pairs correlation for  
$\langle$ $cos(\phi_1)$ $cos(\phi_2)$ $\rangle$ and  
$\langle$ $sin(\phi_1)$ $sin(\phi_2)$ $\rangle$ we do not cause a change in
the results of Ref\cite{mono}. This is because they depend on the difference
between the two terms and thus subtract out.
In this paper we contend that the event average of the azimuthal two particle 
angles $\phi_1$ and $\phi_2$ with respect to the reaction plane as defined above
captures the azimuthal correlation structure of the Au + Au mid-central 
collisions. We use Ref\cite{mono} modified as pointed out above to
generate a event average $\phi_1$ and $\phi_2$ plane distribution. In
Figure 3 we show the event averaged $\phi_1$ and $\phi_2$ plane distribution
for all particle pairs which we have generated from the above method.

In Figure 4 we show the event averaged $\phi_1$ and $\phi_2$ plane distribution
for opposite charged particle pairs. Opposite charges are greatly influenced by
the fragmentation of the charge neutral quark gluon plasma. We see an asymmetry
in the $\phi_1$ and $\phi_2$ plane distribution, because once we choose one
charge the distribution of the other charge is different due to fragmentation
effects. Finally in Figure 5 we show the event averaged $\phi_1$ and $\phi_2$ 
plane distribution for same charged particle pairs. The same charges must have 
a symmetry in the $\phi_1$ and $\phi_2$ plane, since we have Bose symmetry 
particle 1 and 2 are interchangeable.

We are interested in four three particle azimuthal correlations. These 
correlations are defined by the four equations below for Au + Au 
mid-central collisions at $\sqrt{s_{NN}} =$ 200 GeV. The quadrupole
flow from an impact angle defines the orientation Au Au collision and defined
the angle of $v_2$. This $v_2$ angle is charge independent only determined
by geometry. If we choose $\phi_3$ to be charge independent and coming from
particles which range over the whole event, we can replace $\phi_3$ with 
$\psi_2$ the quadrupole flow angle for the first and third equations. For each 
event there will also be a hexapole flow angle for $v_3$ which gives a $\psi_3$ 
which can be used in the second and fourth equations.
\begin{equation}
C_{112} = \langle cos(\phi_1 + \phi_2 - 2\psi_2)\rangle ,
\end{equation}

\begin{equation}
C_{123} = \langle cos(\phi_1 + 2\phi_2 - 3\psi_3)\rangle ,
\end{equation}

\begin{equation}
C_{132} = \langle cos(\phi_1 - 3\phi_2 + 2\psi_2)\rangle ,
\end{equation}

\begin{equation}
C_{143} = \langle cos(\phi_1 - 4\phi_2 + 3\psi_3)\rangle ,
\end{equation}

From above we saw that $\psi_2$ is zero. This requires that all events
are rotated in azimuth such that $\psi_2$ becomes zero. From above we see
that $\psi_3$ points along the $-90^\circ$ axis once we make sure that the
out of plane $v_1$ points along the $90^\circ$ axis.

We will consider particles produced around central rapidity and
concentrate on two particle azimuthal correlations(see above). There is an 
another two particle azimuthal correlation which does not depend on any axis
 
\begin{equation}
\delta =  \langle cos(\phi_1 - \phi_2)\rangle .
\end{equation}

There is a simple relationship between equation 14 and equation 18 since 
$\psi_2$\ = 0.. 

\begin{equation}
C_{112} = \langle cos(\phi_1 + \phi_2) \rangle = \langle cos(\phi_1)cos(\phi_2)
\rangle - \langle sin(\phi_1) sin(\phi_2) \rangle ,
\end{equation}

\begin{equation}
\delta = \langle cos(\phi_1 - \phi_2) \rangle = \langle cos(\phi_1) cos(\phi_2) 
\rangle + \langle sin(\phi_1) sin(\phi_2) \rangle  .
\end{equation}

Thus $\langle$ $cos(\phi_1)$ $cos(\phi_2)$ $\rangle$ and  
$\langle$ $sin(\phi_1)$ $sin(\phi_2)$ $\rangle$ are also values we calculate.

Finally the quadrupole and the hexapole are global event objects while $\delta$
the two particle correlation is a more local effect. In the appendix we show
that just the presents of $v_2$, $v_3$ and a $\delta$ can generate three
particle correlation, but we are dealing with the Case II of the appendix
which consider only particles produced around central rapidity and the
results from Au + Au mid-central collisions at $\sqrt{s_{NN}} =$ 200 GeV are
inconsistent with the appendix.

The two particle azimuthal correlation($\delta$) for all charged particles
is defined by

\begin{equation}
\delta = \langle cos(\phi_1 - \phi_2) \rangle,
\end{equation}

which is independent of the reaction plane. this is because if any angle is 
added to the azimuth it will be subtracted away in the difference. In Figure 6
we plot $\delta$ for all charge particle pairs as a function $\Delta \phi$. 
$\Delta \phi$ is given by

\begin{equation}
\Delta \phi = (\phi_1 - \phi_2).
\end{equation}

Figure 6 show the presence of a dipole and a quadrupole. In Figure 7 we 
plot $\delta$ for unlike sign particle pairs as a function $\Delta \phi$. 
Figure 7 shows the presence of a dipole, a quadrupole and a hexapole. In 
Figure 8 we plot $\delta$ for like sign particle pairs as a function 
$\Delta \phi$. Figure 8 shows the presence of a negative dipole and a 
quadrupole. 

Finally using $\phi_1$ and $\phi_2$ azimuthal plane we calculate $C_{112}$, 
$C_{123}$, $C_{132}$, and $C_{143}$. We also calculate $\langle$ $cos(\phi_1)$ 
$cos(\phi_2)$ $\rangle$ and  $\langle$ $sin(\phi_1)$ $sin(\phi_2)$ $\rangle$. 
The results of these calculations for unlike sign pairs are shown in Table I
and Figure 9, where as the calculations for like sign pairs are shown in 
Table II and Figure 10. These are a prediction of these correlations and the
$\phi_1$ and $\phi_2$ azimuthal plane can be used to summarize
the azimuthal two particle correlations with respect to the reaction plane
for heavy ion Au + Au mid-central collisions.

\section{Acknowledgments}

This research was supported by the U.S. Department of Energy under Contract No.
DE-AC02-98CH10886.

\section{Appendix}

In this appendix we consider the three particle correlators 
$C_{112}$ and $C_{123}$ and how they can be generated from a pure two particle 
correlation by interacting with a $v_2$ and a $v_3$ of the overall system.

The starting point for our discussion is the definition $C_{112}$ and 
$C_{123}$\cite{three}.

\begin{equation}
C_{112} = \langle cos(\phi_1 + \phi_2 - 2\phi_3)\rangle ,
\end{equation}

\begin{equation}
C_{123} = \langle cos(\phi_1 + 2\phi_2 - 3\phi_3)\rangle ,
\end{equation}
where $\phi_1$, $\phi_2$, $\phi_3$ denote the azimuthal angles of the
produced particle 1, produced particle 2 and produced particle 3. 

For the generated events used in this appendix no charge is used even though 
they are charged particles. Thus we are considering charge independent 
correlations. The events generated are modulated by $v_2$ with an axis at 
random plus a $v_3$ modulation which is at a different axis. $v_2$ is a factor 
four larger than the generated $v_3$. In the above equations the third 
particle is considered a reference. Equation 1 we are referring to the $v_2$ 
axis and equation 2 we are referring to the $v_3$ axis. If we want to consider 
referring to the $v_2$ axis in equation 2 we can define another equation.  

\begin{equation}
C_{132} = \langle cos(\phi_1 - 3\phi_2 + 2\phi_3)\rangle .
\end{equation}

Let us consider Case I a system where two particles which are at a given $\eta$
have a two particle correlation given by 

\begin{equation}
\delta =  \langle cos(\phi_1 - \phi_2)\rangle .
\end{equation}

This two particle correlation has no preferred axis. Particle three is at
a different $\eta$ such there is no two particle correlation between it and
the other two particles. However particle three will respond to the over all
$v_2$ and $v_3$, thus being a good reference particle. 

In general the two particle correlation has no axis and would not generate
an effect in equation 1, 2 and 3, but with the modulation of $v_2$ and
$v_3$ a correlation is pick up leading to

\begin{equation}
C_{112} = \delta v_2,
\end{equation}
and

\begin{equation}
C_{123} = \delta v_3
\end{equation}
see Ref\cite{CMS}.
In $C_{112}$ particle 1 and 2 have the two particle correlation which is 
modulated by the $v_2$ and pick up by the second harmonic reference of the
third particle. The same is true for $c_{123}$ where particle 1 and 2 have the 
two particle correlation which is modulated by the $v_3$ and pick up by the 
third harmonic reference of the third particle. These correlations will
vanish if the over all modulations go to zero. The result we get for equation
3 is

\begin{equation}
C_{132} = \delta v_2.
\end{equation}
 
In $C_{132}$ particle 1 and 2 have the two particle correlation which is 
modulated by the $v_2$ and pick up by the second harmonic reference of the
third particle. 

Let us consider Case II a system where all three particles which are at a 
given $\eta$ have a two particle correlation given by 

\begin{equation}
\delta =  \langle cos(\phi_1 - \phi_2)\rangle  =  \langle cos(\phi_1 - \phi_3)\rangle  =  \langle cos(
\phi_2 - \phi_3)\rangle .
\end{equation}

As before

\begin{equation}
C_{112} = \delta v_2.
\end{equation}

This is true no matter which particles are pick for 1, 2 and 3. and by 
symmetry

\begin{equation}
C_{123} = C_{132} = \delta v_2.
\end{equation}

Finally for this case we have

\begin{equation}
C_{112} = C_{123} = C_{132} = \delta v_2.
\end{equation}

\end{document}